\documentstyle[12pt,aaspp4,flushrt]{article}
\def\plotthree#1#2#3{\centering \leavevmode
\epsfxsize=.31\columnwidth \epsfbox{#1} \hfil
\epsfxsize=.31\columnwidth \epsfbox{#2} \hfil
\epsfxsize=.31\columnwidth \epsfbox{#3} \hfil}

\slugcomment{Accepted for Astronomy Letters}

\begin{document}

\title{Scattering of X-ray emission lines by a helium atom}

\author{L.Vainshtein\altaffilmark{1,2}, R.Sunyaev\altaffilmark{2,3}, E.Churazov\altaffilmark{2,3}}
\affil{$^1$ P.N. Lebedev Physical Institute, Moscow, Russia}
\affil{$^2$ MPI fur Astrophysik,
Karl-Schwarzschild-Strasse 1, 85740 Garching, Germany}
\affil{$^3$ Space Research Institute (IKI), Profsouznaya 84/32, Moscow 117810, 
Russia}

\begin{abstract}
	The differential cross section for scattering of the astrophysically
important X-ray emission lines by a helium atom is calculated with an
accuracy sufficient for astrophysical applications. For a helium atom
an energy ``gap'' (due to the structure of the energy levels) is
twice larger than for a hydrogen atom 
and the ``compton profile'' (due to broader distribution of
an electron momentum) is significantly shallower. This opens principle
possibility to distinguish helium and hydrogen contributions,
observing scattered spectra of X-ray emission lines. 
With the appearance of a new generation of X-ray telescopes, combining
a large effective area and an excellent energy resolution it may be
possible to measure helium abundance in the molecular clouds in the
Galactic Center region, in the vicinity of AGNs or on the surface of
cold flaring stars. 
\end{abstract}

\section{Introduction}
	 The spectra of
X-ray emission lines, scattered by neutral atoms, contain information on the
atoms themselves. This effect was used to study the electron momentum
distribution in atoms by observing scattered spectra of X--ray and gamma--ray
lines (e.g. \cite{ep70}). As pointed out by Sunyaev and Churazov
(1996) in astrophysical 
conditions it is not unusual, when X--ray emission lines (e.g. iron fluorescent
$K_\alpha$ line at 6.4 keV) are scattered by neutral matter.
Of particular importance are the smearing of the Compton
backscattering peak (for the lines in the X-ray band) due to the motion of
electrons bound in neutral 
atoms and the energy ``gaps'' below the initial line energy due to the
structure of energy levels of the discrete states. It was noted that
these effects 
may provide principle possibility to distinguish helium and hydrogen
contributions to the scattered spectrum and thus to measure helium
abundance in the scattering media. Below we calculate cross sections
for the 
scattering of X--ray photons by a helium atom with an accuracy
sufficient for astrophysical applications. Details of the scattering
of X--ray lines by a hydrogen molecule will be described elsewhere. 

\section{Calculations}
	The differential cross section for a scattering of an X--ray photon
(for a given change of direction and energy of the photon) by a light atom 
is given by the expression (e.g. \cite{hei,ep70}): 
\begin{eqnarray}        \label{basic1}
\frac{d\sigma}{d\Omega dh\nu}=\left(\frac{e^2}{mc^2}\right)^2 \left(\frac{\nu_2}{\nu_1}\right) (\vec{e_1}\vec{e_2})^2
\sum_f | \langle f | e^{i\vec{\chi}\vec{r}} | i \rangle |^2 \times
\delta(\Delta E_{if} - \Delta h\nu) 
\end{eqnarray} 
where indices $1$ and $2$ refer to the photon before and after scattering,
$\nu$ is the photon frequency, $\vec{e}$ is the direction of polarization, $i$
and $f$ denote initial and final states of the electron,
$\vec{\chi}=(\vec{k_1}-\vec{k_2})/\hbar$, $\vec{k_1}$ and $\vec{k_2}$
are the initial and final momenta of the photon. In the most important
astrophysical applications the initial state $i$ of the 
electron corresponds to the ground state of an atom. 
According to the type of the final state of the electron (after scattering)
the process is 
conventionally  subdivided into three channels: Rayleigh scattering
($f\equiv i$), 
Raman scattering ($f$ corresponds to one of the exited discrete states
of the atom) and  Compton scattering ($f$ corresponds to the continuum
state, i.e. photon ionizes atom). 
For the hydrogen atom all calculations can be done analytically, but 
for more complex systems (e.g. helium) one has to perform numerical
integration using approximate wave--functions. Note that exactly the
same matrix element as in (\ref{basic1}) appears in 
the expression of the cross section  for the scattering of fast
electrons in the Born 
approximation (\cite{llqm}).   
We therefore used code ``ATOM'' (\cite{atom}) primarily intended for
the 
calculation of the excitation or ionization cross sections of atoms
and ions by  an 
electron impact. For the calculation of the matrix element in (\ref{basic1}) it
uses one--electron wave functions obtained in a potential that provides
experimental values of discrete atomic levels. The accuracy of this method 
should be sufficient  for typical astrophysical applications.

The following contributions to the sum over $f$ were calculated: all
transitions to the discrete states with the same spin ($\Delta 
S=0$) and the principle quantum number $n\le 4$; transitions to
continuum (i.e. Compton scattering) to all states with the orbital quantum
number $l\le 12$. 

The results of our calculations are represented as function
$H(\Delta h\nu,q)=\sum_f | \langle f | e^{i\vec{\chi}\vec{r}} | i \rangle
|^2 \times \delta(\Delta E_{if} - \Delta h\nu)$, where $\Delta h\nu$ is the
change of photon energy 
during the scattering in eV and $q=|\vec{\chi}|a_0$ is a change of the
photon momentum in atomic units 
($\frac{\hbar}{a_0}$, where $a_0$ is the Bohr radius). The cross section was
calculated for the values of $q$ from $10^{-3}$ to $6$, thus covering
the possible range of momentum changes for astrophysically important
lines with energies below 10 keV. In the case of Compton scattering
$H(\Delta h\nu,q)$ was normalized per 1 eV. In order 
to calculate the scattered spectrum of an unpolarized monochromatic
line (by one bound electron) one should 
multiply $H(\Delta h\nu,q)$ by the factor
$\left(\frac{e^2}{mc^2}\right)^2\times
0.5\times(1+cos^2\theta)\times\left(\frac{\nu_2}{\nu_1}\right)$. I.e. 
\begin{eqnarray}        \label{use}
\frac{d\sigma}{d\Omega dh\nu}=\left(\frac{e^2}{mc^2}\right)^2
\left(\frac{\nu_2}{\nu_1}\right) 0.5\times(1+cos^2\theta) \times H(\Delta
h\nu,q) 
\end{eqnarray} 
where $\theta$ is the scattering angle, $\Delta h\nu = h\nu_1 -h\nu_2$ (in eV), $q=
((\nu_2 sin\theta)^2+(\nu_2 cos\theta - \nu_1)^2
)^{1/2}\times\frac{a_0}{c}$. 
The tables are available as FITS files at
http://hea.iki.rssi.ru/\verb1~1chur/compton/helium.fits and 
http://hea.iki.rssi.ru/\verb1~1chur/compton/hydrogen.fits. 
A subsample of the data is given in Table 1. Note that for the elastic
scattering (which corresponds to zero change of photon energy) an 
increase of the cross section due to coherent scattering was taken into
account. 

\tiny
\begin{table}[t]
\caption{Subsample of the table for helium. The 
values of momentum transfer are in atomic units. 
The first two columns are the change
of photon energy (in eV) and type of the process ( 0 -- Rayleigh, 1
-- Raman, 2 -- Compton). The other elements in the table are values of
$H(\Delta h\nu,q)$.} 
\begin{center}
\begin{tabular}{lcllllll}\hline
& & \multicolumn{5}{c}{$q$, atomic units} \\
$\Delta E$, eV & Process & 1.01E--03 &  4.53E--03 &  2.03E--02 &  9.09E--02 & ...&    1.82E+00 \\
\hline
      0.000& 0 &  2.00E+00  &  2.00E+00 &   2.00E+00 &   1.99E+00 &   ...&    6.61E--01 \\
     20.607& 1 &  4.30E--03 &  4.31E--03 &  4.31E--03 &  4.48E--03 &  ...&    1.99E--02 \\
     21.169& 1 &  1.01E--07 &  2.04E--06 &  4.09E--05 &  8.13E--04 &  ...&    8.61E--03 \\
     22.911& 1 &  1.02E--03 &  1.02E--03 &  1.02E--03 &  1.06E--03 &   ...&   5.26E--03 \\
     23.064& 1 &  2.45E--08 &  4.93E--07 &  9.90E--06 &  1.97E--04 & ...&     2.93E--03 \\
     23.065& 1 &  3.08E--15 &  1.24E--12 &  5.01E--10 &  1.98E--07 & ...&     1.09E--04 \\
     23.664& 1 &  3.97E--04 &  3.97E--04 &  3.97E--04 &  4.10E--04 & ...&     2.11E--03 \\
     23.726& 1 &  9.67E--09 &  1.94E--07 &  3.89E--06 &  7.76E--05 & ...&     1.29E--03 \\
     23.727& 1 &  1.53E--15 &  6.20E--13 &  2.50E--10 &  9.89E--08 & ...&     6.35E--05 \\
     23.728& 1 &  4.31E--23 &  3.49E--19 &  2.82E--15 &  2.22E--11 & ...&     5.74E--07 \\
     24.790& 2 &  2.08E--08 &  4.19E--07 &  8.41E--06 &  1.68E--04 &  ...&    7.04E--03 \\
     24.799& 2 &  2.08E--08 &  4.18E--07 &  8.41E--06 &  1.68E--04 & ...&     7.04E--03 \\
     24.809& 2 &  2.08E--08 &  4.18E--07 &  8.40E--06 &  1.68E--04 &  ...&    7.04E--03 \\
     24.820& 2 &  2.08E--08 &  4.17E--07 &  8.39E--06 &  1.67E--04 &  ...&    7.05E--03 \\
     24.830& 2 &  2.07E--08 &  4.17E--07 &  8.38E--06 &  1.67E--04 &  ...&    7.05E--03 \\
     24.841& 2 &  2.07E--08 &  4.16E--07 &  8.37E--06 &  1.67E--04 &  ...&    7.05E--03 \\
     24.853& 2 &  2.07E--08 &  4.16E--07 &  8.36E--06 &  1.67E--04 &  ...&    7.05E--03 \\
     24.865& 2 &  2.07E--08 &  4.15E--07 &  8.35E--06 &  1.67E--04 &  ...&    7.06E--03 \\
     24.878& 2 &  2.06E--08 &  4.15E--07 &  8.33E--06 &  1.66E--04 &  ...&    7.06E--03 \\
       ...    & ..&  ...         &  ...         &  ...         &  ... & ...&  ... \\
    479.046& 2 &  4.92E--13 &  9.90E--12 &  1.98E--10 &  4.00E--09 &  ...&    3.17E--06 \\
    499.167& 2 &  4.17E--13 &  8.38E--12 &  1.68E--10 &  3.38E--09 &  ...&    2.62E--06 \\
    520.178& 2 &  3.53E--13 &  7.09E--12 &  1.42E--10 &  2.86E--09 &  ...&    2.18E--06 \\
    542.120& 2 &  3.00E--13 &  6.02E--12 &  1.20E--10 &  2.43E--09 &  ...&    1.80E--06 \\
    565.034& 2 &  2.53E--13 &  5.06E--12 &  1.01E--10 &  2.04E--09 &  ...&    1.49E--06 \\
    588.962& 2 &  2.13E--13 &  4.26E--12 &  8.56E--11 &  1.72E--09 &  ...&    1.23E--06 \\
    613.950& 2 &  1.80E--13 &  3.59E--12 &  7.22E--11 &  1.45E--09 &  ...&    1.02E--06 \\
    640.043& 2 &  1.52E--13 &  3.03E--12 &  6.09E--11 &  1.22E--09 &  ...&    8.45E--07 \\
    667.292& 2 &  1.28E--13 &  2.55E--12 &  5.12E--11 &  1.03E--09 &  ...&    6.97E--07 \\
    695.748& 2 &  1.15E--13 &  2.29E--12 &  4.61E--11 &  9.28E--10 &  ...&    6.08E--07 \\
    725.463& 2 &  9.08E--14 &  1.80E--12 &  3.62E--11 &  7.29E--10 &  ...&    4.79E--07 \\
    756.494& 2 &  7.72E--14 &  1.53E--12 &  3.08E--11 &  6.19E--10 &  ...&    3.97E--07 \\
\hline
\end{tabular}
\end{center}
\end{table}
\normalsize

\begin{figure}[t]
\plottwo{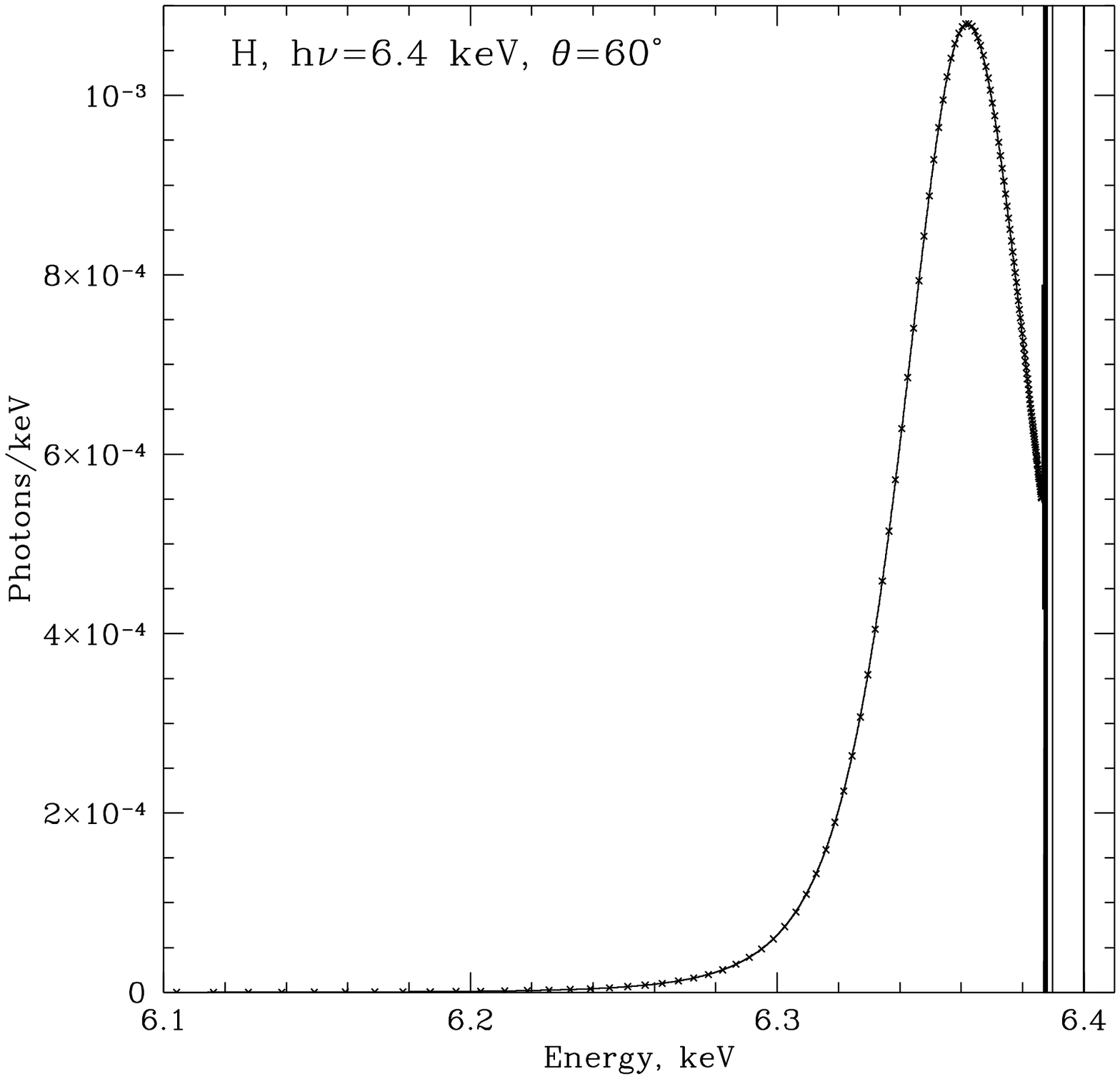}{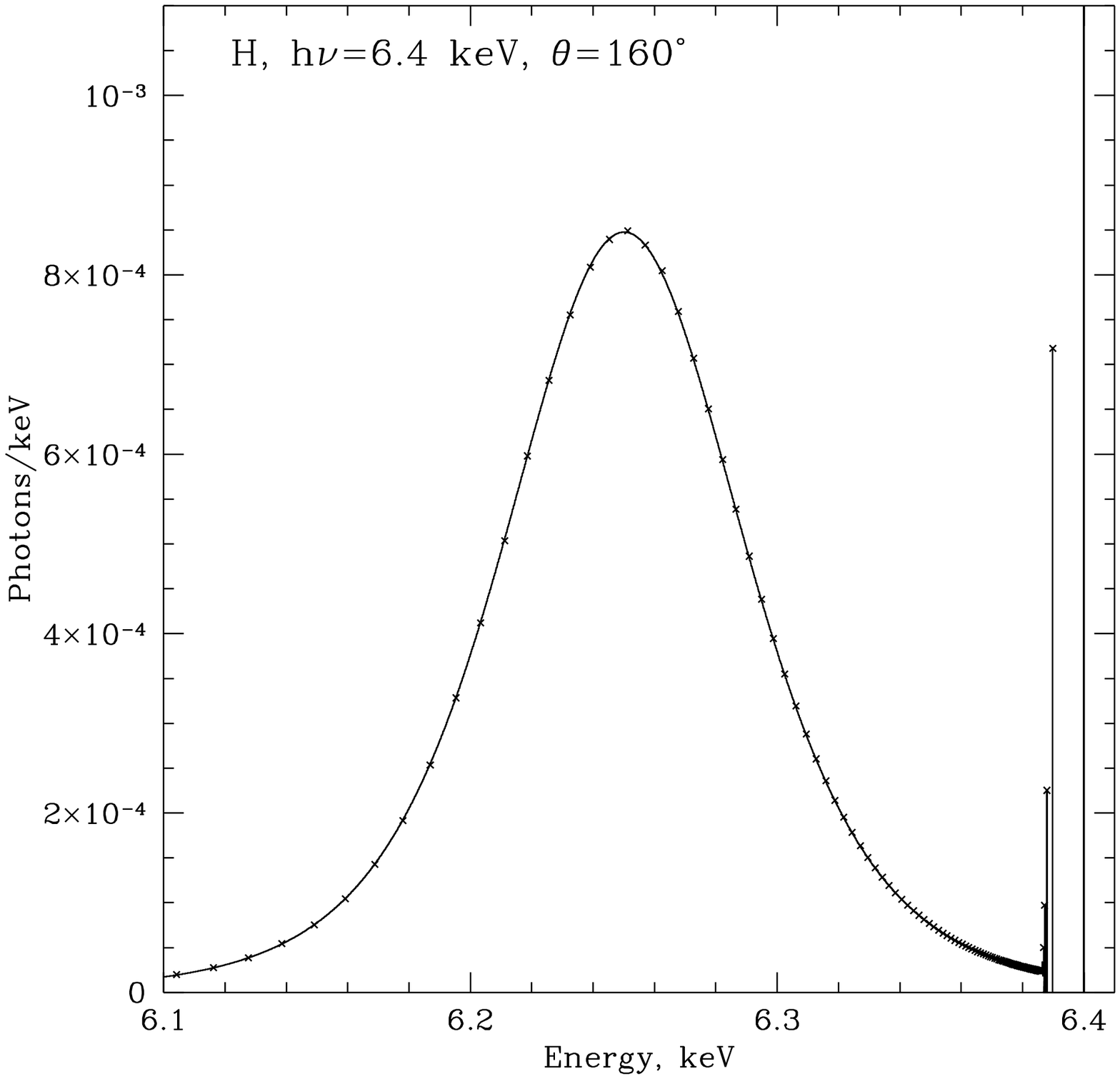}
\caption{
Comparison of ATOM results and analytical calculations for
hydrogen. The solid line corresponds to analytical formulae. Crosses
mark ATOM results. } 
\label{comphy}
\end{figure}

For the hydrogen atom all calculations can be performed
analytically (see e.g. \cite{ep70,llqm}). Shown in Fig.\ref{comphy}
are the scattered 
spectra of the monochromatic 6.4 keV line for scattering angles of $60^\circ$
and $160^\circ$. The solid line shows analytical calculations, while crosses
mark ATOM results. 

\begin{figure}[t]
\plottwo{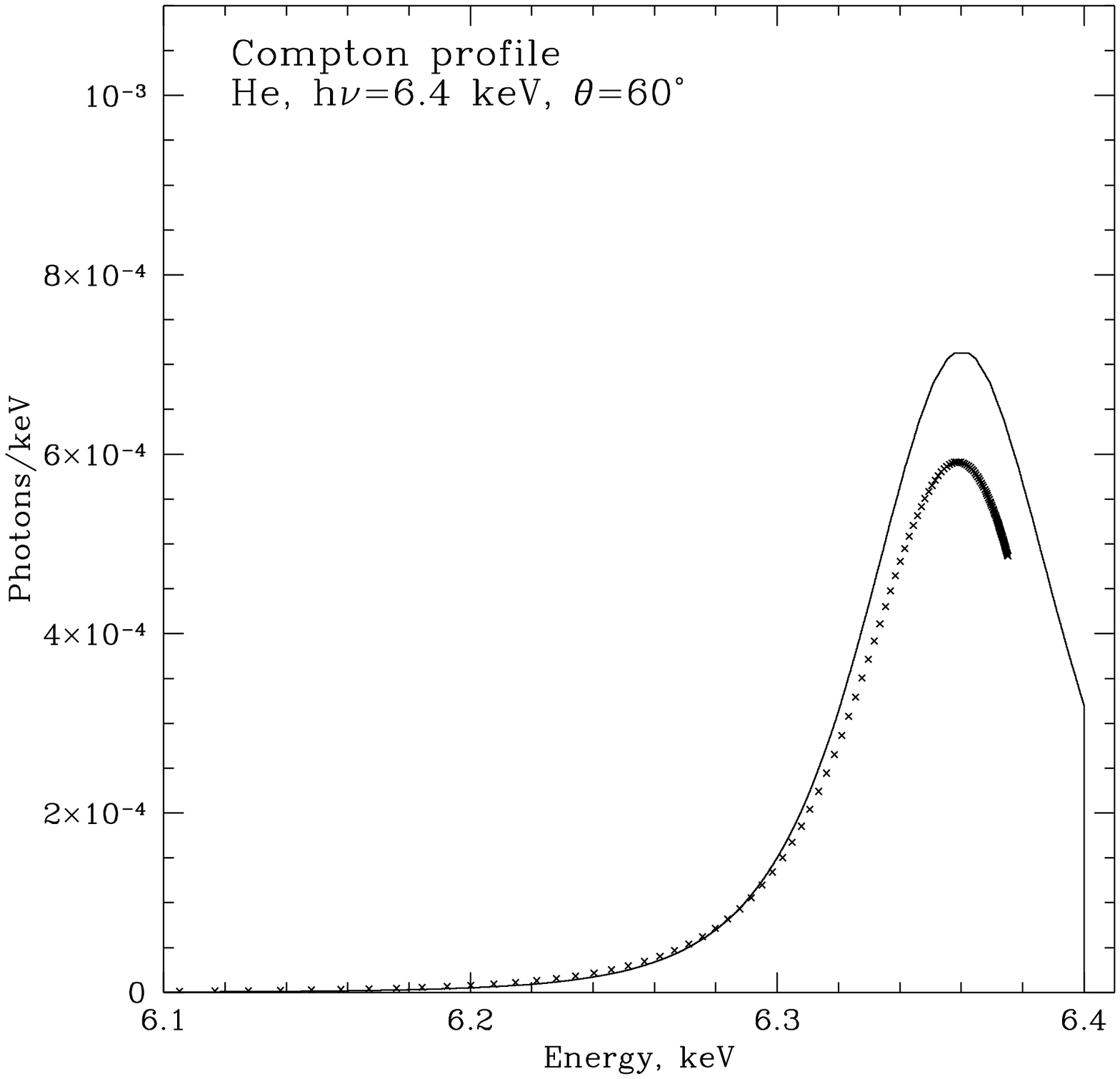}{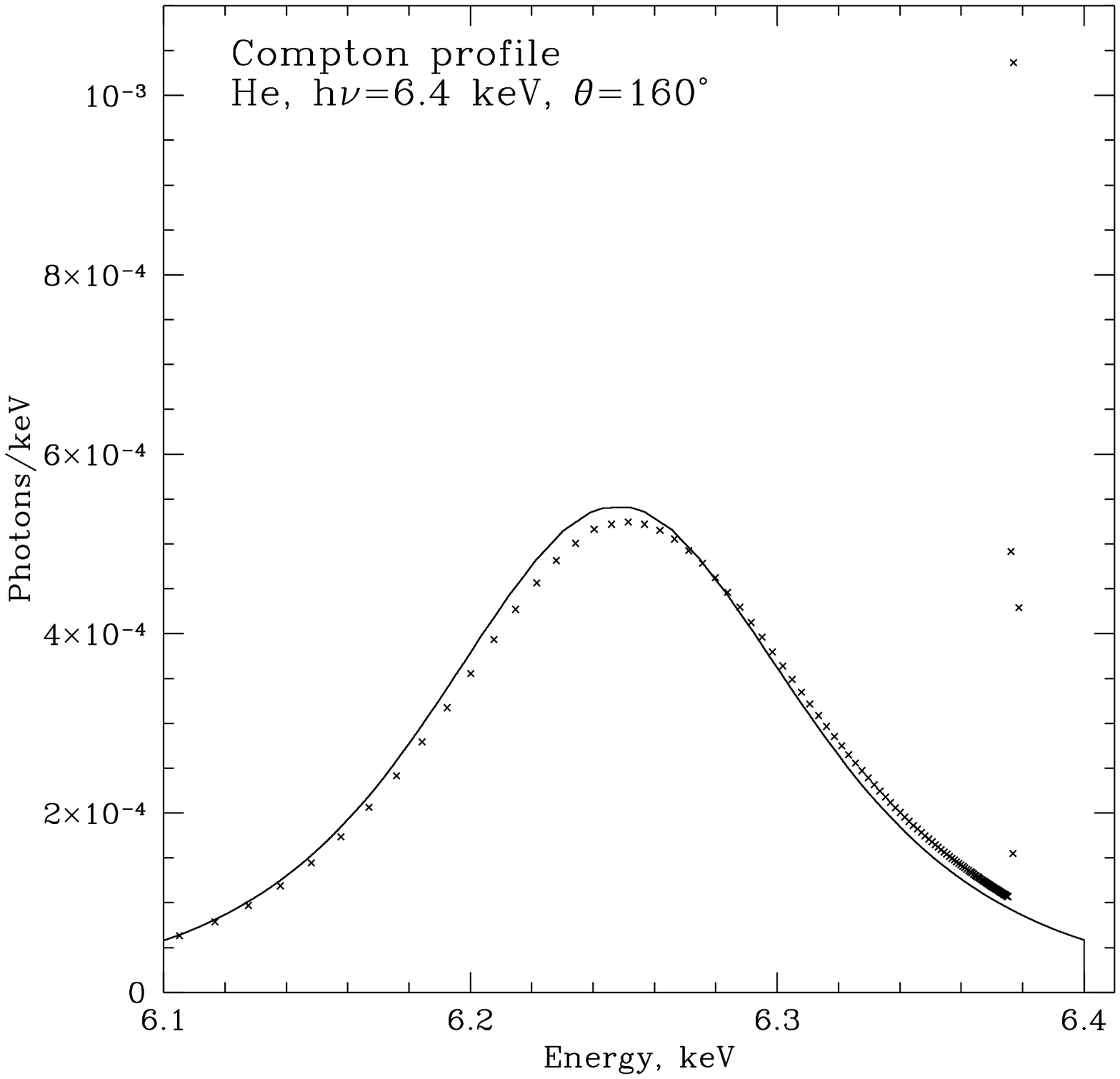}
\caption{
Comparison of ATOM results and impulse approximation for helium. For the left
figure the points corresponding to transitions to discrete levels (ATOM
results) are beyond the range of the plot.
} 
\label{comphe}
\end{figure}

For the helium atom calculations of the Compton profile in the so called
``impulse approximation'' are available (e.g. \cite{e70}). This
approximation is valid  as long as energy transferred from a photon to
an electron sufficiently exceeds the binding energy of the electron in
the atom. Shown in Fig.\ref{comphe} are the 
spectra of the monochromatic 6.4 keV line scattered by a helium
atom. The solid line shows the ``impulse approximation'' and crosses mark
the ATOM results. Note that the impulse approximation curve was artificially
extended in the region where the change of photon energy is small. One can see
that calculations in the impulse approximation approaches ATOM results
in the region where this approximation is applicable.

\section{Scattering by a helium atom}
\begin{figure}[t]
\plotthree{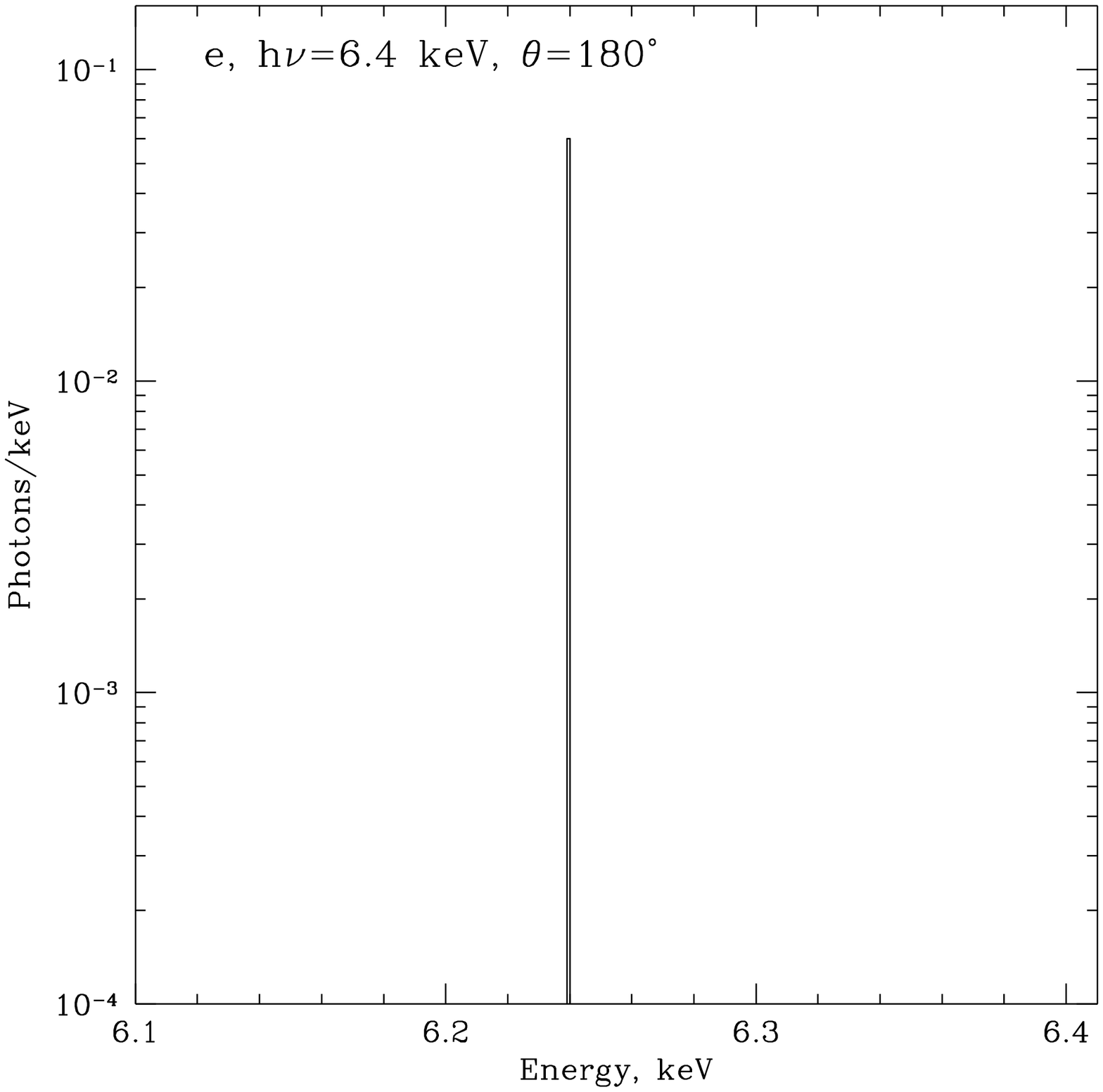}{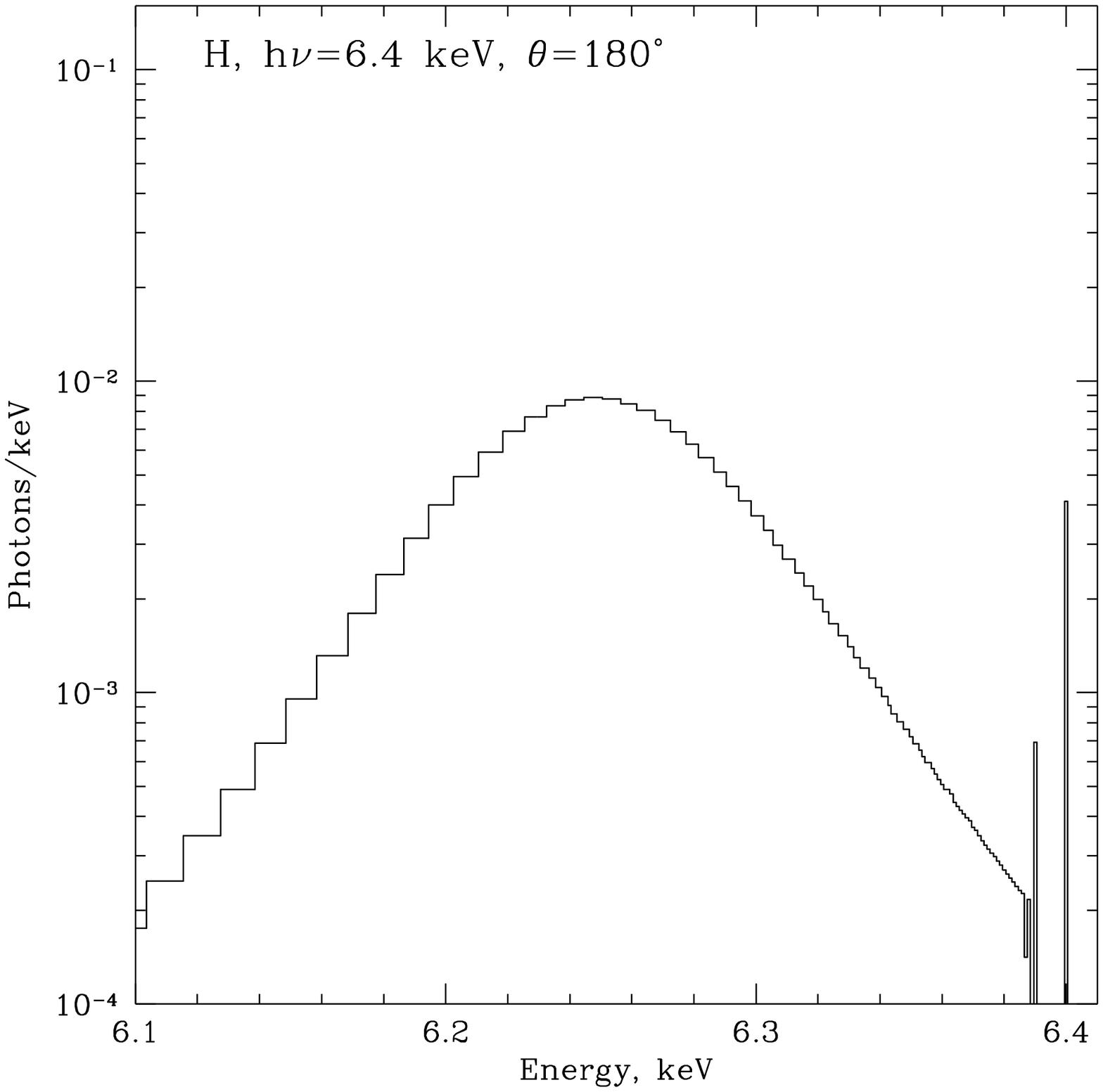}{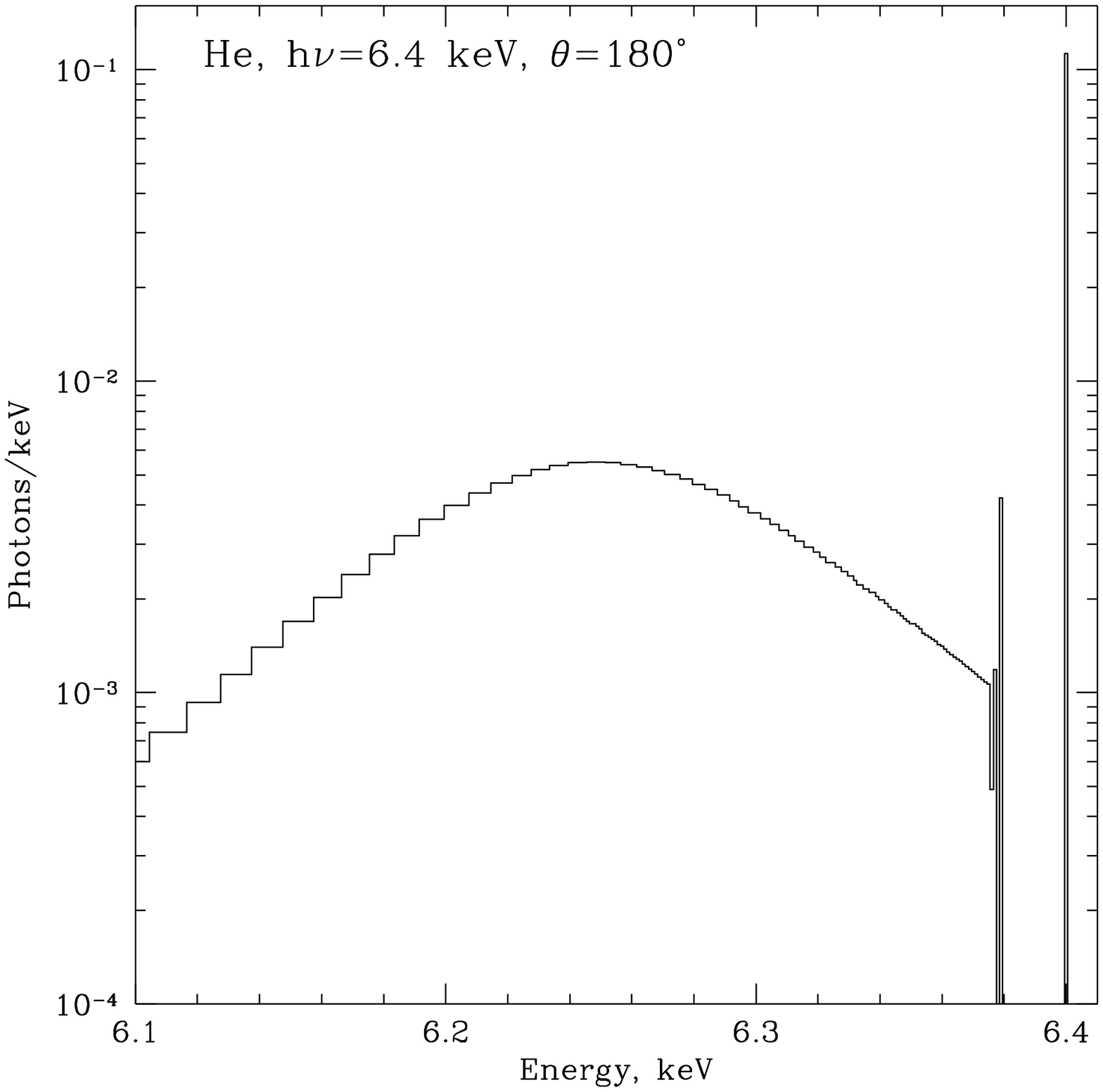}
\caption{
Spectra of the 6.4 keV photons scattered by 180 degrees by free
electron at rest (left), hydrogen atom (middle) and helium atom
(right). 
} 
\label{spang}
\end{figure}

      Shown in Fig.\ref{spang} are the spectra of the monochromatic 6.4
keV line scattered by 180 degrees by a free electron at rest, hydrogen
and helium atom. For a free electron the energy of the photon after
scattering is unambiguously related to the scattering angle:
$\nu_2=\nu_1 (1 - \frac{h\nu_1}{m_ec^2} cos\theta)$. Motion of
electron in atom causes broadening of the energy distribution due to
the Doppler effect. Features close to the initial line energy are related
to the excitation of discrete levels of an atom. For a helium atom (in
comparison with scattering by a hydrogen atom) the different structure 
of the energy levels and the larger binding energy of the electrons cause
deviations in the scattered spectra. The role of the elastic scattering is
higher for the helium atom since (i) larger excitation energy means that
larger energy transfer is required to excite or ionize helium atom and
Rayleigh scattering will dominate till larger scattering angles, (ii)
Rayleigh scattering is ``enhanced'' in helium due to the coherence effect.
For Raman scattering the most apparent difference is the energy
``gap'' (between the unshifted line and first Raman satellite) for
a helium atom, which is twice as large as for a hydrogen atom. This is the
most important effect from the point of view of potential use of recoil
profiles for the determination of helium abundance. For Compton scattering
the larger binding energy of the electrons in a helium atom means a
broader momentum distribution of the electron in the ground state and
therefore a broader recoil profile. As noted by Sunyaev and Churazov
(1996) this broadening is 
completely analogous to the thermal broadening of the line profiles, being
scattered by ``warm'' free electrons. Larger binding energy just means
that the 
effective ``temperature'' of the electrons in helium atoms is a factor
of  $\sim$ 2 higher than that in hydrogen atoms.

\section{Simplest astrophysical applications}
\begin{figure}[t]
\plotone{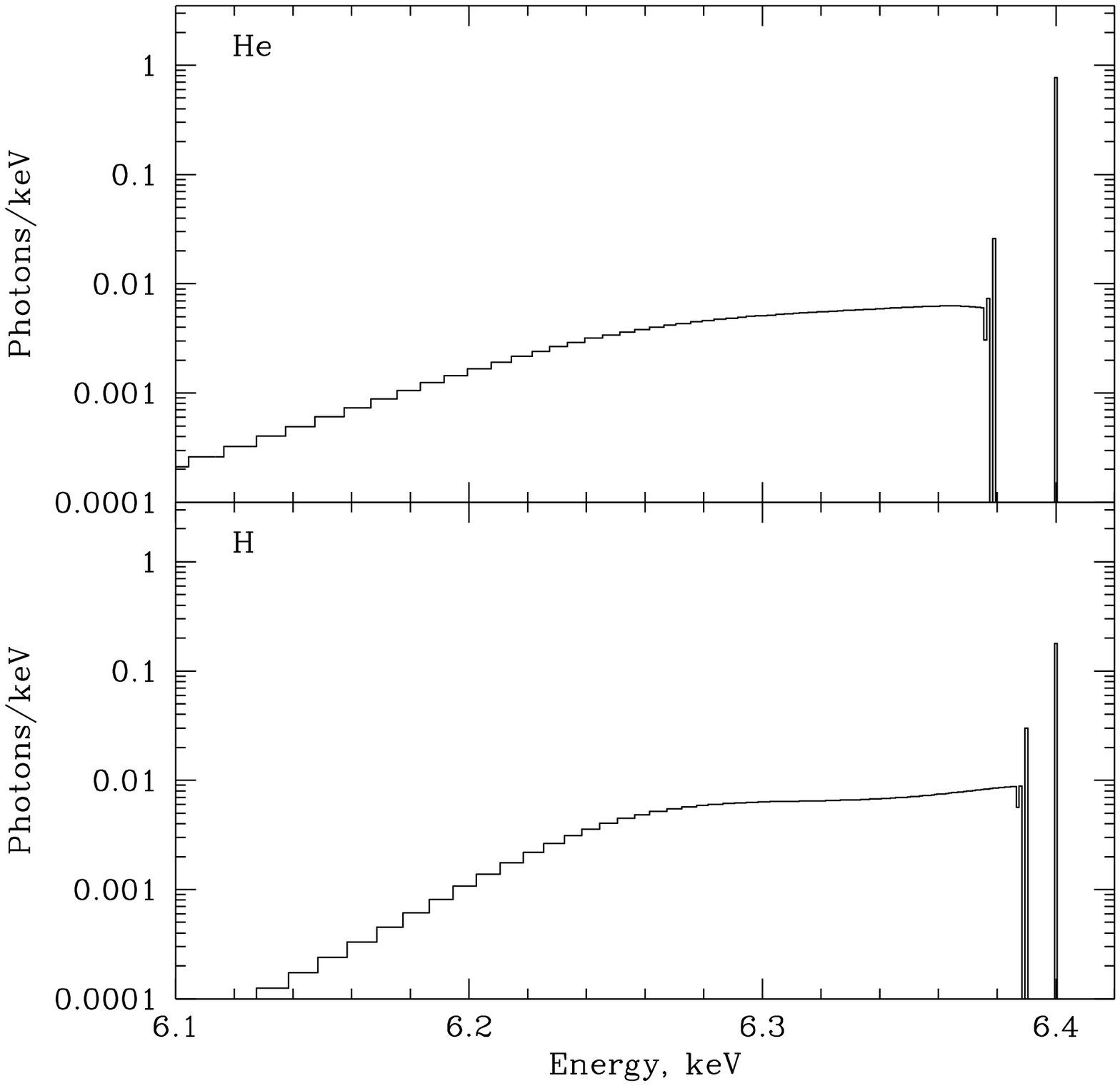}
\caption{
Spectra of monochromatic 6.4 keV line, scattered by hydrogen and helium
atoms and averaged over all angles. One can expect to see such spectra
if the source of the monochromatic line is located at the center of
an optically thin cloud.} 
\label{av}
\end{figure}

Detailed modeling of the possible astrophysical applications
is beyond the scope of this article. We just mention below some basic
problems associated with practical implementation of the possibility
of measuring the helium abundance in cosmic objects using X--ray data. One
of the principle difficulties is associated with the fact that usually
scattered line flux is only a fraction ($\sim \tau_T=N_H\times R \times
\sigma_T$ ) of the direct (unscattered) line flux. In turn the 
contribution of photons scattered by helium atoms is also a fraction
($\sim 0.2$ for canonical Solar abundance) of the scattered
flux. Furthermore, really distinct feature related to scattering by 
helium atoms is a $\sim 20$ eV gap below the unshifted line. 
Obviously detection of such a feature requires very high sensitivity
and excellent energy resolution, which may be provided by future planned
X--ray mission like HTXS (Constellation) (\cite{whi97}) and Xeus
(\cite{tur97}). 

        One of the most natural situations, when an X--ray emission line is
scattered by neutral media is related to the neutral iron fluorescent
$K_\alpha$ line at 6.4 keV. This line always appears, when neutral matter
(e.g. molecular cloud) is illuminated by X--rays with an energy above 7.1 keV.
Scattering of the $K_\alpha$ photons by neutral hydrogen or helium in the same
cloud naturally produces a ``recoil'' wing at the low energy side of
the line. The fraction of photons in the wing is equal to the Thomson depth
of the cloud (in the optically thin case). The actual shape of the wing depends
on the angular distribution of scattering matter and primary radiation,
which may be different in each particular situation. However one
can get an idea of it's shape by considering simplest case of the point source
of $K_\alpha$ line, located at the center of the optically thin cloud. In
this case the recoil wing corresponds to the scattered spectrum (\ref{basic1})
averaged over all angles (see Fig.\ref{av}). 

The $K_\alpha$ line itself consists of two components
($K_{\alpha_1}$ and $K_{\alpha_2}$) with energies of 6.404 and 6.391 keV and
relative intensities 2:1. According to experimental data,
the intrinsic widths of these lines are $\sim$ 2.65 and 3.2 eV respectively
(e.g. \cite{sl76}). The substructure of these components, their
energies and widths can be affected by the type of a chemical bond
i.e. they may differ for an isolated iron atom and an
atom in dust grains. We note that such a ``chemical'' shift of
$K_\alpha$ or $K_\beta$ fluorescent lines (which perhaps may
be at the level of $\le$ eV; see e.g. \cite{gm97} for the chemical shift of
$K_\beta$ line of Cr) could open an interesting possibility to
distinguish fluorescence due to free atoms or dust, provided sub eV
energy resolution of the spectrometers. Anyway the presence of two components
($K_{\alpha_1}$ and $K_{\alpha_2}$) separated by $\sim$ 13 eV and 
especially low energy wing of the Lorentzian line profile will
significantly complicate the detection of features associated with
scattering by helium atoms.

For the $K_\alpha$ lines of lower $Z$ elements (compared to iron) the
situation with energy resolution is somewhat better, since the intrinsic width
and separation of the components will decrease, while the energy gap
of $\sim$20 eV will remain unchanged. However lower $Z$ elements have
lower fluorescent yield (see e.g. \cite{bam72}), lower reflectivity
from the neutral media (which scales as $\sigma_T/\sigma_{ph}(E)$) and
larger fraction of elastic scattering. 

Emission lines of the thermal plasma can also be of interest
e.g. considering spectra of the Solar flares reflected by the
Solar photosphere. The most favorable conditions would exist if
one can spatially resolve the flare itself and scattering region.
In this case the wings of the direct (unscattered) line will not
contaminate the spectrum. The Doppler widths of the lines should not
be too serious 
problem since for the plasma at e.g. $T_e=2~10^7$ K  the Doppler FWHM
of the iron line at $\sim$ 6 keV is about 3 eV and the wings decline
very rapidly.  Low energy lines again have the advantage of smaller widths
of features with respect to the gap width and disadvantage of smaller
reflectivity and increased coherent scattering fraction.

\section{Conclusions}
 The differential cross section for scattering of the astrophysically
important X-ray emission lines by  a 
helium atom is calculated with the accuracy sufficient for astrophysical
applications. The differences in the width of the Compton backscattering
peak and energy gap, due to discrete energy levels, open the principle
possibility to distinguish helium and hydrogen contributions observing
the scattered spectra of sharp features. However practical
implementation demands extremely high sensitivity and energy
resolution. As far as observations of the energy gap is concerned, the  major
difficulty is associated with the contamination of that region (just 20 eV
below the line) by the Lorentzian wings of the unscattered (or
elastically scattered) line itself. Depending on the future instruments
characteristics, fluorescent or emission lines of different elements
may become favorable for such studies.

This work was supported in part by RBRF grants 96-02-18588 and 97-02-16919.


\end{document}